\documentclass[a4paper]{article}
\usepackage{graphicx}
\usepackage[utf8]{inputenc}
\usepackage{fullpage}
\usepackage[english]{babel}
\usepackage{csquotes}
\usepackage{siunitx}
\usepackage{arydshln}
\usepackage{float}
\usepackage{subfigure}
\usepackage{verbatim}

\title{Identifying Geographic Clusters:\\ A Network Analytic Approach}

\author{Roberto Catini\thanks{LIME Research Unit, IMT Institute for Advanced Studies, Lucca, Italy, email: roberto.catini@imtlucca.it}, Dmytro Karamshuk\thanks{Department of Informatics, King's College, London, UK, email: dmytro.karamshuk@kcl.ac.uk},\\ Orion Penner\thanks{LIME research unit, IMT Institute for Advanced Studies, Lucca, Italy, email: orion.penner@imtlucca.it}, Massimo Riccaboni\thanks{LIME research unit, IMT Institute for Advanced Studies, Lucca, Italy and KU Leuven, Leuven, Belgium, email: m.riccaboni@imtlucca.it}}

\date{}
\begin{document}

\maketitle
\begin{abstract}
\noindent In recent years there has been a growing interest in the role of networks and clusters in the global economy. Despite being a popular research topic in economics, sociology and urban studies, geographical clustering of human activity has often studied been by means of predetermined geographical units such as administrative divisions and metropolitan areas. This approach is intrinsically time invariant and it does not allow one to differentiate between different activities. Our goal in this paper is to present a new methodology for identifying clusters, that can be applied to different empirical settings. We use a graph approach based on k-shell decomposition to analyze world biomedical research clusters based on PubMed scientific publications. We identify research institutions and locate their activities in geographical clusters. Leading areas of scientific production and their top performing research institutions are consistently identified at different geographic scales.
\end{abstract}


\section{Introduction}

Recent advancements in our understanding of the evolution of scientific and technological systems stem from the theoretical conceptualization, and empirical investigation, of the dynamic relationship between the notions of innovation cluster and network. The idea of agglomeration -- or local concentration -- of related activities is hardly new, it was indeed raised by Alfred Marshall in his seminal work~\cite{marshallprinciples} and subsequently developed by other scholars such as Jacob~\cite{jacobs1970economy} and Perroux~\cite{perroux1950economic}, to name a few. The revival of the cluster idea among economic geographers, sociologists and economists of innovation, follows from the influential work of Porter~\cite{1990competitive} and the global fame of the Silicon Valley~\cite{saxenian94}. Scholars have contributed a multiplicity of interpretations of the original concept, thus resulting in a certain degree of theoretical and empirical confusion~\cite{martin2003deconstructing,maskell2006,maskell2007}. Since the notion of cluster is only vaguely defined in terms of geographical scale and network properties, the cluster idea has been operationalized in different ways. The  lack of consensus on a precise definition of clusters compounded with difficulties in identifying the location of inventive activities has led many economists to proxy innovation clusters with administrative regions.\footnote{By the same token, the tendency to analyze the size of cities by means of administrative urban areas has been recently criticized~\cite{rozenfeld2009area}.} Two elements are core in most cluster definitions. First, clusters are constituted by related activities. Second, clusters are geographically proximate groups of interlinked individuals and organizations. This common ground notwithstanding, boundaries of network communities and geographical areas are not clearly identified. Therefore, most of the available evidence has used standard industrial classification systems and administrative regions or zip codes to investigate localized knowledge spillovers and agglomeration economies. Apart from a few recent contributions~\cite{ter2011co,balland2013proximity,delgado2013}, by taking regional and industrial boundaries as fixed and pre-determined, most of empirical studies falls short in describing the evolution and plasticity of clusters both in geographical and technological terms. Regional clusters of innovative activities depend on the networks that arise from linkages among co-located inventors. Learning by interacting is a crucial force pulling inventors into clusters and a crucial factor of success for innovative clusters. Therefore, the local presence of bundles of related knowledge bases is constantly evolving. As a consequence, clusters change their location, size and performances over time and tend not to fit static administrative boundaries~\cite{cerina2013networks}.

As noted elsewhere, a full theory of clusters should answer to the questions on ``what, how, why as well as when, where and who'', focusing in particular on causal relationships (the ``why'' question) \cite{maskell2006,press2006}. To achieve these goals quality data is required, as are analytical tools to study the evolution of clusters \cite{delgado2013}, i.e. the emergence of new clusters, how they grow and decline, and how they shift into new fields \cite{menzel2010,boschma2006}. To analyze cluster life cycles data should be: (a) referring to specific activities; (b) geo-localized at the finest possible level of resolution (i.e. street address); (c) time resolved to know when exactly a given activity has been performed; and (d) individual and institution names should be disambiguated to properly identify the economic agents who have been active in the cluster over time. The increasing availability of geo-referenced micro-data for the implementation of a bottom-up approach for clusterint related activities in space and time. There is a rapidly growing literature in geography \cite{feser2000,marcon2003,marcon2009}, spatial statistics \cite{diggle1991} and economics \cite{duranton2005}, that has advanced the identification of clusters of various phenomena using a series of spatial point-process and distance-based techniques. More recently, new and rigorous methods have been developed to identify bundles of related activities \cite{delgado2013,delgado2012}. Our work complements previous contributions in this field by developing a bottom-up network approach to map and trace the evolution of clusters over time.
Most of the literature so far used zip code centroids to analyze the spatial distribution of activities based on geographical concentration measures \footnote{One exception being \cite{feser2000} in which street addresses are also considered but only for a limited number of records.}.

In this paper we introduce a new method based on network-analytic tools instead of spatial density measures. Our approach takes into account the relational patterns among co-located activities. Moreover, we develop a computationally efficient solution and apply it to a large database of street-level records. We focus on the geographical loci of scientific production, as opposed to economic production or human population density. Thus, because agglomeration patterns for R\&D and production differ, we used the geographic distributions of scientific output instead of more conventional variables such as industry employment, plant output, or population statistics. For innovation clusters, patents and scientific publications are typically used in cluster identification because scientific and technological outputs tend to correspond to the locations in which inventors innovate. As in Furman et al. (2005), we use scientific publications to locate biomedical research activities~\cite{furman2005public}. Specifically we use the PubMed database of approximately 23 million bibliographic records corresponding to peer-review publications in the biomedical sciences. To avoid relying on predetermined administrative boundaries, such as NUTS3 regions or Metropolitan Statistical Areas (MSAs), and to allow for the examination of the innovation landscape {\it within} cities, we employ and develop a variety of novel methods. To observe the geography of biomedical research activity on the finest scale possible we have exploited existing geocoding APIs to arrive at high resolution (street level) coordinates for a significant portion of all biomedical publications. To street level geocoded data we apply an aggregation algorithm drawing inspiration from previous complex networks research. This approach works in a straightforward manner: treating all geocoded papers as nodes, we connect nodes (papers) that are less than a certain distance apart. This is repeated until there are no further nodes within the certain distance to link to. At the completion of this algorithm we arrive at a set of large geographic clusters, akin to MSAs, that capture the bulk of biomedical production in a geographic area. To better understand the institutional make up of the clusters produced by the algorithm we develop an (institutional) disambiguation algorithm that allows for the identification of institutions from noisy addresses. The high resolution geocoding also allows for deeper analysis of the internal structure of  clusters. Specifically, we carry out a ``k-core'' analysis in which low degree nodes are recursively removed from the graph, exposing the internal structure of a cluster.

A similar approach has been recently adopted by Alc{\'a}cer and Zhao (2012) who use a clustering identification algorithm to analyze the role of internal cluster relationships in the global semiconductor industry \cite{alcacer2012local}. Our method differs from the one they implemented in four aspects. First, we use online mapping services (such as Yahoo or Google) to localize researchers. The same approach can be applied to extract location information from address fields in any data source including patents, publications, research projects and firm locations. Second, our method does not critically depend of the choice of parameters, such as the neighborhood radius and the contour threshold. Our method works at a finer level of resolution, based on the intuitive notion of walking distance between inventors. Third, apart from localizing the inventor, our algorithmic approach also provides a disambiguated name of the affiliation of the inventor. Thus, we are able to position firms and research institutions in clusters. Fourth, we defined clusters by the actual distribution of inventor locations following a spatially-embedded graph approach. By doing that, our method applies for any definition of local graph links. While we use a specific empirical context (global biomedical research) for illustrative purposes, the insights and methodologies are general and our methodology can be applied in other empirical settings. Our approach is particularly valuable in all cases in which information is available about the relationships between individuals and/or firms. In those cases, our methodology can take as an input the geo-referenced network of collaborations in which only a subset of co-localized nodes are connected.

The manuscript is organized in the following manner. First we carefully explain the geocoding and affiliation disambiguation approaches that generate the extremely high resolution geographical and institutional data required for our analysis. Then we describe in full the graph based algorithm we employ to identify the relevant clusters and cores of biomedical scientific production around the world. Following that we present the results of our method, showing the three largest clusters produced by our algorithm and probing their internal core structure with complex network analysis techniques and then showing the structural change of the San Francisco cluster resulting from a particular development thereby illustrating the potentiality of our approach in analyzing the temporal evolution of clusters. We follow with general observations and trends arising from our analysis and conclude by discussing the avenues of future enquiry opened by our approach.

\section{Preparing affiliation and location data}

To capture and map knowledge production in the biomedical research field, we develop a novel approach to collect and cluster publicly-available knowledge in the life sciences. Namely, we draw on the National Library of Medicine’s PubMed database. PubMed contains bibliographic records of over 23 million publications in the biomedical sciences, with full coverage dating back to 1966. The main advantage of using PubMed data is that it is open access and a number of other high value data sets have already been extracted from it, for example the Authority dataset that provides fully disambiguated author names \cite{torvik2005probabilistic}. An additional benefit of using of an open access dataset is it allows other scholars to verify and extend our methodology.  Moreover, Pubmed provides a detailed classification of the content of publications, based on the Unified Medical Language System (UMLS). The Pubmed dataset does not, however, provide extensive information about the location at which the research was carried out: only the affiliation and address for each paper’s corresponding author is provided. While practices vary across the disciplines and sub-disciplines of biomedical science it is generally the case that the corresponding author is either the junior scientist that carried out the majority of the research, or the senior scientist (Principle Investigator) in charge of the research group in which the research was carried out. In either instance, in the majority of cases it seems a reasonable proxy that the corresponding author's affiliation corresponds to the location at which the majority of work for any given publication took place.

We emphasize that the methods described below may be applied to a wide variety of data sets, as long as there is a sufficiently descriptive address associated with each piece of data. The criteria for sufficiently descriptive differs slightly between the geocoding approach and the affiliation disambiguation algorithm. In the case of geocoding, the address must contain information beyond simply a city name. A full street address, for example, is certainly descriptive enough to result in a high resolution geocoding. Similarly, department name, or university name, or company name (with a city name for large multinationals) are generally enough to produce a high quality geocoding. In the case of the affiliation disambiguation algorithm the address (or other data in the record) must contain some hints as to the institution of origin, be it university, department, company, {\it etc.} To be concrete, examples of data to which both approaches may be applied include Thomson Reuters Web of Science and Elsevier Scopus publication databases, the European Patent Office (EPO) patent database and the Orbis global firm database by Bureau Van Dijk. On the other hand, an example of data ill suited for both approaches is United States Patent Office (USPTO) inventor address data, in which only the city name and state are included.

\subsection{PubMed data geocoding}

Despite the lack of structured geographic information in the PubMed dataset (the only available ``country'' field is empty for the majority of publication records), missing details on authors' locations can be effectively mined from affiliation strings. Provided in free text format, affiliation strings usually include information about authors' organizations and their detailed addresses. To map this textual data to a structured geographic information we exploit an out-of-box solution provided by the Yahoo Geocoding API. The API is provided as a web-service which for an input affiliation string returns, in JSON format, a list of potential location matching. Each location record includes, among others, a pair of latitude and longitude coordinates and a level of confidence (called ``quality'') that the suggested location is correct. In contrast, administrative information (e.g., country, region, city) is often absent or inaccurate in the API responses. Therefore, we manually map latitude and longitude pairs associated with each affiliation string to geographic administrative areas using GADM dataset.\footnote{Global Administrative Areas, http://gadm.org/} As a result, each publication record in our dataset is assigned one or several location tags with geographic coordinates and corresponding administrative information. However, it is worth noting that in the analysis presented later in this paper we use only the geographical coordinates of each publication, making no use of the administrative unit information.

Despite the fact that the proposed geocoding technique can effectively handle unstructured affiliation information from the PubMed dataset, the output of the geocoder requires some cleaning before it can be used for further analysis. Firstly, we discard all PubMed records with empty affiliation strings. Further, the geocoded data is filtered by the ``quality'' value associated with each location match. Particularly, we omit all location records with accuracy coarser than the street level. In Table~\ref{tab:statistics} we report the resulting statistics. We note, that the filtered dataset accounts for the vast majority ($63$\%) of publications in the original dataset. In this table ``distinct locations'' refer to unique [longitude, latitude] coordinates returned by the geocoding at which at least one publication exists. This is less than the number of papers because many papers get aggregated into the same coordinate, for example because they come from the exact same department. We note that we observe significant variation country to country in the rate of accurate geocoding. For the USA we see 84.5\%, Japan 76.3\%, the United Kingdom 70.0\%, all quite solid. But on the other hand several countries are much lower, Italy and Korea coming in below 35\% and Russia much below that. In cases where we observe poor geocoding translation and transliteration appear to be the main bottlenecks and future improvements to the approach are required.

\begin{table}[ht]
	\centering
	{\small
	\begin{tabular}{lr}
		\hline
		Number of geocoded publications & 9,754,647\\
		Number of publications with ``accurate geocoding'' & 6,163,880 \\
		Number of distinct locations& 149,951 \\
		Number of distinct locations with ``accurate geocoding'' & 130,738 \\
		\hline
	\end{tabular}
	}
	\caption{General statistics on the geolocation. In this context, a geocoding is said to be accurate if its quality is better than the city level.}
	\label{tab:statistics}
\end{table}

\subsection{Disambiguation of authors' affiliations}

To characterize the spectrum of research activities concentrated in a given geographic region we need to systematize information about authors' affiliations in our data set. The major challenge here lies in extracting organization names from a free form text which constitutes the affiliation field in the PubMed data set and, further, identifying organizations from the extracted names. This problem, typically called `disambiguation', is non-trivial due to a number of reasons which, among others, include various spelling discrepancies, different ordering of words, use of abbreviation etc. In this current work, we extract and disambiguate organization names from affiliation strings by applying the approach proposed in~\cite{jonnalagadda2010nemo}. We justify our choice by the fact that the method proposed by Jonnalagadda {\it et al.} (2010) performs reasonably well when applied to the PubMed data set whilst not requiring any information sources other than the affiliation strings. At this point in time it is worthwhile pointing out that there has been extensive previous work on author name disambiguation in publication databases~\cite{torvik2005probabilistic, NameDisambigISI} and patent databases~\cite{lai2011disambiguation}. While fundamental goals of affiliation disambiguation and author disambiguation are similar, the techniques that need to be employed in each case are rather different. In the affiliation case the primary challenge is the diversity of unique strings that can refer to the same institution and there is minimal metadata available. In the author disambiguation case a lot of metadata is available and there is minimal variation in strings ({\it i.e.} John Smith may appear as J. Smith or J.M. Smith) but the major challenge is determining which J. Smith papers belong to which J. Smith.


This algorithm proceeds in two steps. First, it extracts the part of the affiliation string corresponding to organization name. This is achieved by classifying words and, further, phrases composed of words in three different classes of information: address, institution and department information. To classify individual words we exploit a set of dictionaries, including a thesaurus of geographic names, organization keywords and person names. Once classes are assigned to individual words, the algorithm proceeds to classifying phrases (we split a string to phrases by commas) where a set of decision rules is applied. For example, a phrase is defined as address information if it contains no organization keywords but a geographic name possibly mixed with numbers (e.g., building number, zip code). In contrast, we say that a phrase defines an institution name if it consists of an organization keyword or an abbreviation mixed with a geographic or a person name, but not a number. For example, in the following affiliation string \emph{``Department of Chemical and Biochemical Engineering, The Advanced Centre for Biochemical Engineering, University College London, Torrington Place, London WC1E 7JE, UK''} the algorithm is able to recognize ``Torrington Place'', ``London WC1E 7JE'', ``UK'' as address-related phrases and ``Department of Chemical and Biochemical Engineering'', ``The Advanced Centre for Biochemical Engineering'', ``University College London'' as organization description phrases. To distinguish between department and institution name we use a heuristic assumption that the one containing geographic or a person name is most probably an institution name, e.g., ``University College London'', and otherwise -- name of a department, e.g.,  ``Department of Chemical and Biochemical Engineering'', ``The Advanced Centre for Biochemical Engineering''. Manual validation of the results suggests that $95$\% of institution names can be extracted correctly.  

In the second step, the algorithm identifies various organization names that correspond to a single institution. The algorithm can spot and tolerate a reasonable number of discrepancies in the way organization names are written, including various spelling mistakes, missing words, different ordering of the words etc. To this aim, we firstly tokenize all previously extracted institution names and build a dictionary of disambiguated words. This allows us to capture insignificant spelling differences between similar words, e.g., plural endings, missing apostrophes, missing letters etc. Particularly, we identify and unify the words that are not more than $n$ characters different from each other, where $n$ is defined as $80$\% of the word length. Further, we compare the phrases by computing the Levenshtein distance defined on the words level: we assume that each word in a phrase is a symbol and define the cost of a word removal as a length of the word and the cost of word replacement as a summary length of both words. We say that two phrases are similar if the Levenshtein distance between them is not larger than $4$. As a result, we get the list of ``normalized'' institution names one of which assigned to each publication record. We note here, that the manual validation of the disambiguation results revealed a high level of accuracy, i.e., $72$\% of institution names were disambiguated correctly.   

\section{The clustering algorithm}


To uncover the clusters underlying urban biomedical research productivity we start at the micro level, employing a geographical aggregation algorithm.
Our algorithm is inspired by the City Clustering Algorithm (CCA) originally introduced by Rozenfeld {\it et al.} (2008) to construct cities without the use of administrative subdivisions. As originally implemented it exploited the population distribution, typically obtained from census data \cite{rozenfeld2008laws}.

Two versions of the CCA algorithm have been developed: a discrete one designed on coarse grained data (e.g., population index defined on a homogeneous grid representing the region of interest) \cite{rozenfeld2009area} and a continuous one, which considers two points on the map as neighbours if the distance between them is below a critical distance $\ell$.
In our case we will employ the continuous version because we are using the geocoded affiliations, and are not constrained by any coarse-graining procedure. 

The actual clustering procedure is performed by repeating the following steps until all points are assigned to clusters:
\begin{itemize}
\item take one arbitrary location and assign it to a new cluster;
\item find all locations closer than $\ell$ to the previous point and assign them to the same cluster;
\item recursively add locations closer than $\ell$ to at least one location already in the cluster until there are no new locations within $\ell$ of any added location.
\end{itemize}

\begin{figure}[H]
	\centering
	\includegraphics[scale=0.45]{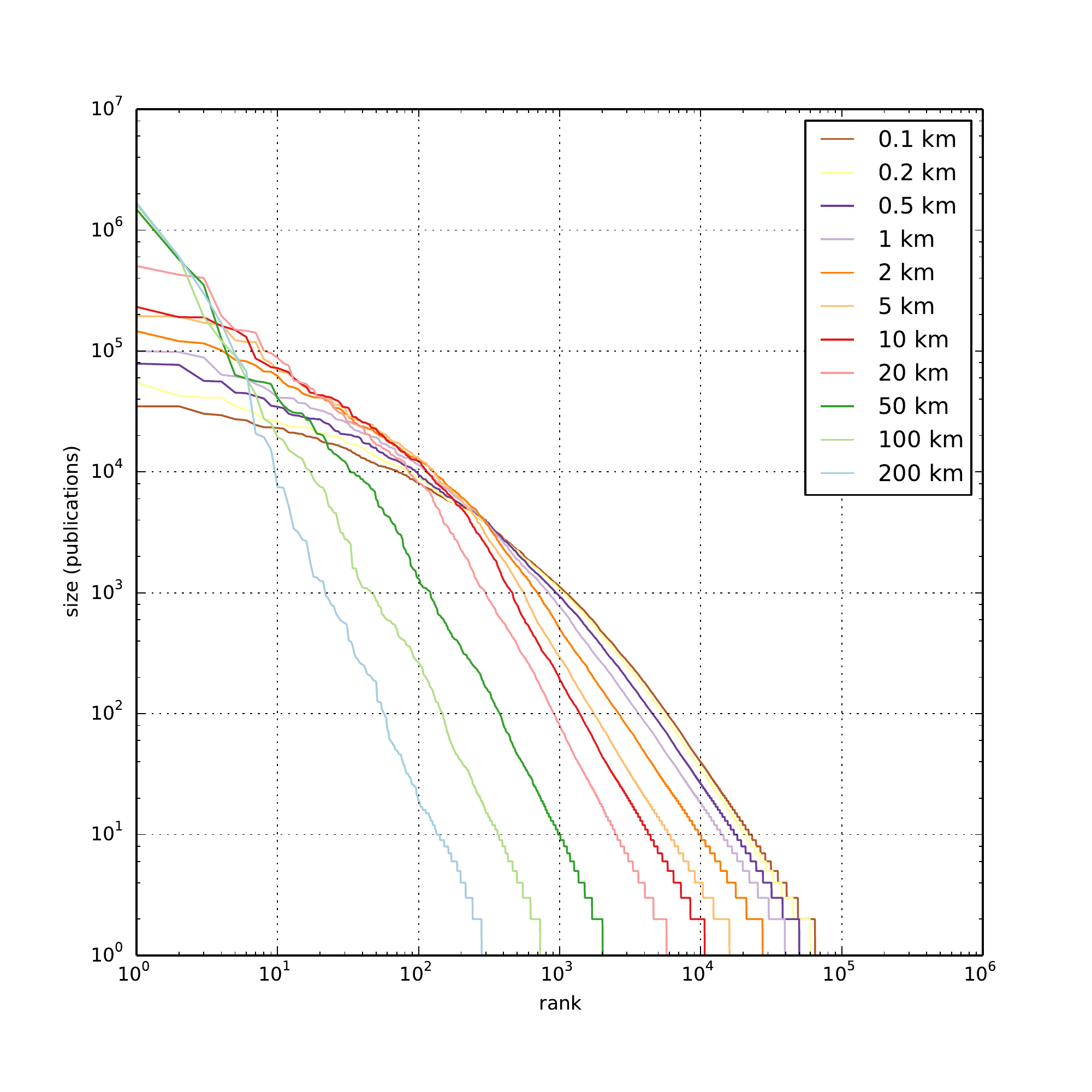}
	\caption{The dependence of the size of a cluster on its rank. For all the distances a power-law behavior is exhibited in the tail.}
	\label{fig:rank-size}
\end{figure}

We have run the algorithm on a wide range of $\ell$ values to select the best value for defining clusters; the results can be found in Figure \ref{fig:rank-size}.

A reasonable distance $\ell$ has been chosen heuristically as \SI{1}{\km}, and approximately corresponds to a ``walking distance''.  We use the Authority data set \cite{torvik2005probabilistic} and our geolocated data to verify if it's really the case that scientists collaborate more if they are at walking distance from each-other. First, we assigned to each author an ``home address'' computing the most frequent location appearing in all his/her publications. Then we extracted from the PubMed data a sample of 1 million publications and then took all the possible pairs of coauthors (i.e. we extracted all the possible 2-combinations without repetition, for each publication with at least two authors). For each pair of authors, we then computed their geographical distance according to their home addresses.

\begin{figure}
\centering
\includegraphics[width=\linewidth]{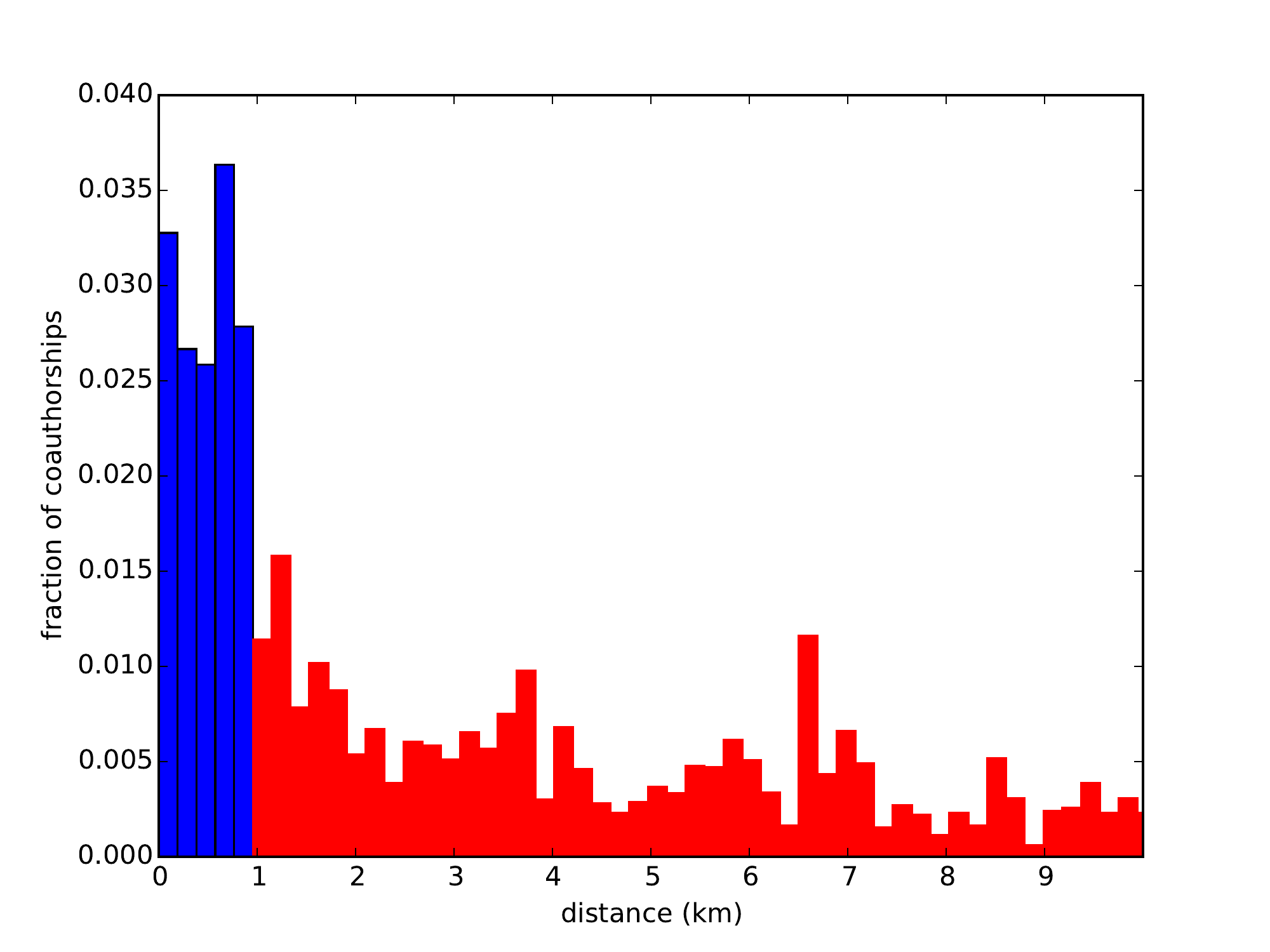}
\caption{Frequency of collaboration observed at a given distance.}
\label{fig:collaborations-distance}
\end{figure}

The results of this analysis is plotted in Figure \ref{fig:collaborations-distance} \footnote{the collaborations at zero distance are discarded, since they represent a very large fraction of the total and would have impeded the readability of the chart}. Figure \ref{fig:collaborations-distance} shows that 1km is a very reasonable length scale for cluster generation as a sharp decrease in collaboration is observed at that distance.

To test the robustness of this distance we have repeated the complete analysis for \SI{0.5}{\km} and \SI{2}{\km}, obtaining similar results; however this choice is by no means definitive and will be the subject of further analysis.
In the Appendix we report the maps of the main biomedical clusters in the US, Europe and Japan. A complete zoomable world map of all biomedical clusters is available at http://goo.gl/X4YUyG. 

\begin{figure}[H]
	\centering
	\includegraphics[width=.65\linewidth]{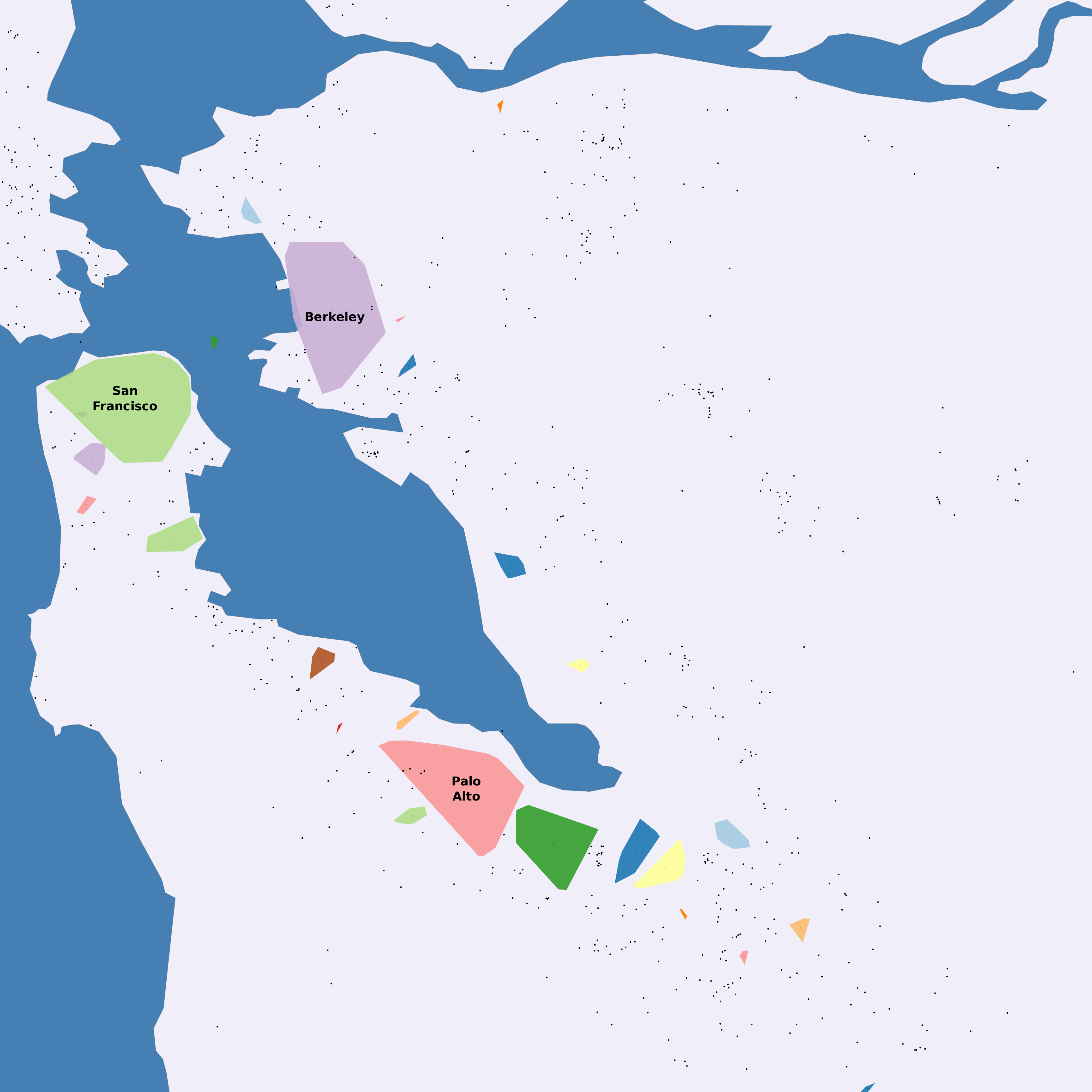}
	\caption{The result of the CCA in the San Francisco Bay Area, for $\ell = \SI{1}{\km}$. Only clusters with at least a hundred publications are shown as colored polygons, other locations are plotted as black dots. Note that with this critical distance the stretch of sea between San Francisco and Berkeley segments the two high productive areas in separate clusters.}
	\label{fig:bay-area}
\end{figure}

In this way clusters of publications can be detected independently from the choice of the starting points, and for the most part they correspond to cities. Note that the correspondence between these components and the administrative definitions (e.g.\ the metropolitan statistical areas of the U.S.) is not bijective because some publications may be isolated geographically due to physical features such as the coastline and will be considered as a different cluster (see the San Francisco case in Figure \ref{fig:bay-area}).

Since a worldwide system that defines regional boundaries does not exists, we have carried out a comparison of our clustering strategy with the following traditional administrative subdivisions: Metropolitan Statistical Areas and FIPS county codes (United States), and NUTS at level 3 (European Union).

To perform the comparison each publication has been matched to the appropriate administrative region and to the corresponding connected component in our graph. Then, for each region, the biggest intersection (in terms of number of publications) with a network component has been selected. For the Metropolitan Statistical Areas (MSAs) each of our components intersects at most one (see Table \ref{tab:MSA}). The fraction of papers in the intersection it is almost unitary (e.g. in the Boston--Cambridge case) there is a bijective correspondence. In the other cases our algorithm has detected multiple components (typically two main ones), and the fraction is approximately half: for example, in Washington--Arlington--Alexandria case, Bethesda is in a component different from downtown Washington. In these cases which is the ``right'' clustering can be debated, but we note that our method has the advantage of using actual research output instead of the distribution of the population.

\begin{table}
\centering
\begin{tabular}{r|rr}
 && fraction in\\
Metropolitan Statistical Area & Publications&biggest intersection \\
\hline
Boston--Cambridge--Newton &           219885 &    91\%  \\
New York--Newark--Jersey City &       150321 &    54\%  \\
Washington--Arlington--Alexandria &   108146 &    55\%  \\
Philadelphia--Camden--Wilmington &    94834  &    68\%  \\
Ann Arbor &                           87113 &     93\%  \\
Houston--The Woodlands--Sugar Land &  74165 &     72\%  \\
Seattle--Tacoma--Bellevue &           64067 &     86\%  \\
Los Angeles--Long Beach--Anaheim &    63718 &     42\%  \\
San Francisco--Oakland--Hayward &     53716 &     45\%  \\
Baltimore--Columbia--Towson &         52281 &     52\%  \\
\hline
\end{tabular}
\caption{Comparison with MSAs. The list is sorted by the size of the matching component and truncated at the 10th entry.}
\label{tab:MSA}
\end{table}

When FIPS county codes and NUTS-3 levels are considered, our method has the advantage not to split clusters that span multiple administrative area, such as Boston-Cambridge in the US and London in Europe.

\section{k-Shell decomposition}

To further understand the complex structure of innovation clusters generated in the previous step we construct the graph of all publications, in which an edge is drawn between two vertices if the distance between them is less than $\ell$.\footnote{Alternative definitions for local connectivity can be implemented as well. For instance, one can draw a link if the technological and geographical distance are both below a given threshold. Another possibility is to consider co-authorship links when the address field is available for all co-authors.} Note that the connected components emerging in this graph correspond to the clusters identified by the algorithm described in the previous paragraph.

The cores of the different components can be revealed using the k-shell decomposition, a method which has been previously applied in complex networks to study the topology of the Internet \cite{carmi2007model}. The idea is to use a percolative process to assign each node in the graph (which in our case is a publication) a measure of its connectivity.

The actual method starts by assigning all isolated nodes a zero index.
Then all nodes with exactly one neighbur (i.e.\ no other publication can be found within $\ell$) are assigned the index 1 and removed from the graph. This step is repeated until no other nodes with one neighbor remain in the graph.
At each further step the index $k$ is increased by one and the nodes with degree $k$ are removed iteratively, until the graph is empty.

Naturally, there is a maximum shell index $k_{max}$ such that once all shells with $k \le k_{max}$ are removed, the graph is empty.
In our case $k_{max}$ can be interpreted as the size (in term of the number of papers) of the area of most intense biomedical research, because it is in practice determined by the total production of very tightly spaced buildings.

Looking at the dynamics of this k-shell process between $0$ and $k_{max}$ two main types of removal steps can be clearly distinguished:

\begin{itemize}
\item a vertex at the geographical frontier is dropped because the low density of papers in the neighborhood is lower than required to be in the k-shell; in this case two parts of the same city become disconnected.
\item a significant number of vertices are dropped at once with a sharp reduction in the fraction of publications remaining in the graph; in this case the vertices are typically positioned very close to each other and the corresponding geographical region becomes empty after the next removal step.
\end{itemize}

\begin{figure}[H]
\centering
\begin{minipage}{.45\linewidth}
  \centering
  \includegraphics[width=.90\linewidth]{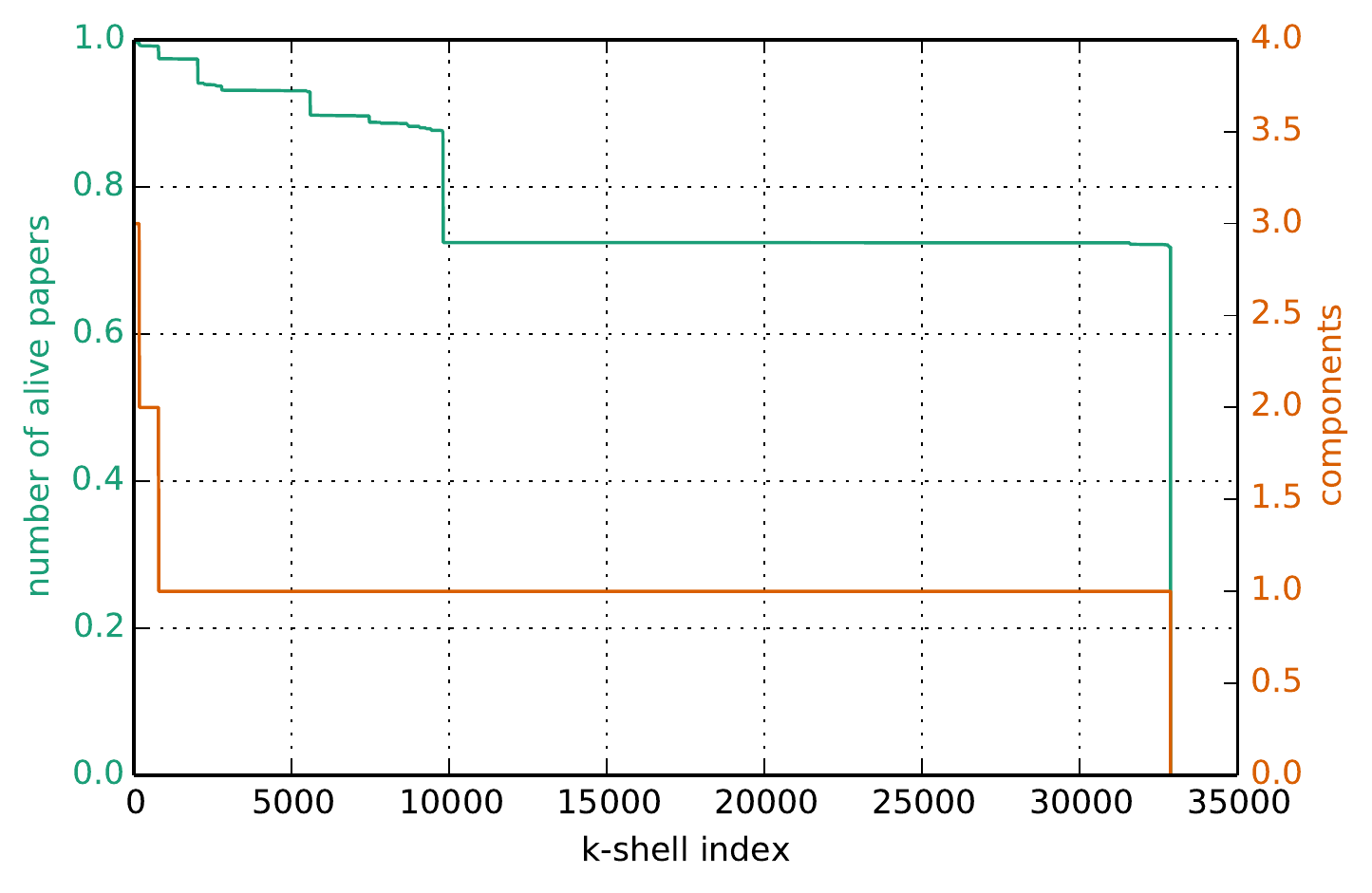}
  \caption{Toronto}
\end{minipage} 
\begin{minipage}{.45\linewidth}
  \centering
  \includegraphics[width=.90\linewidth]{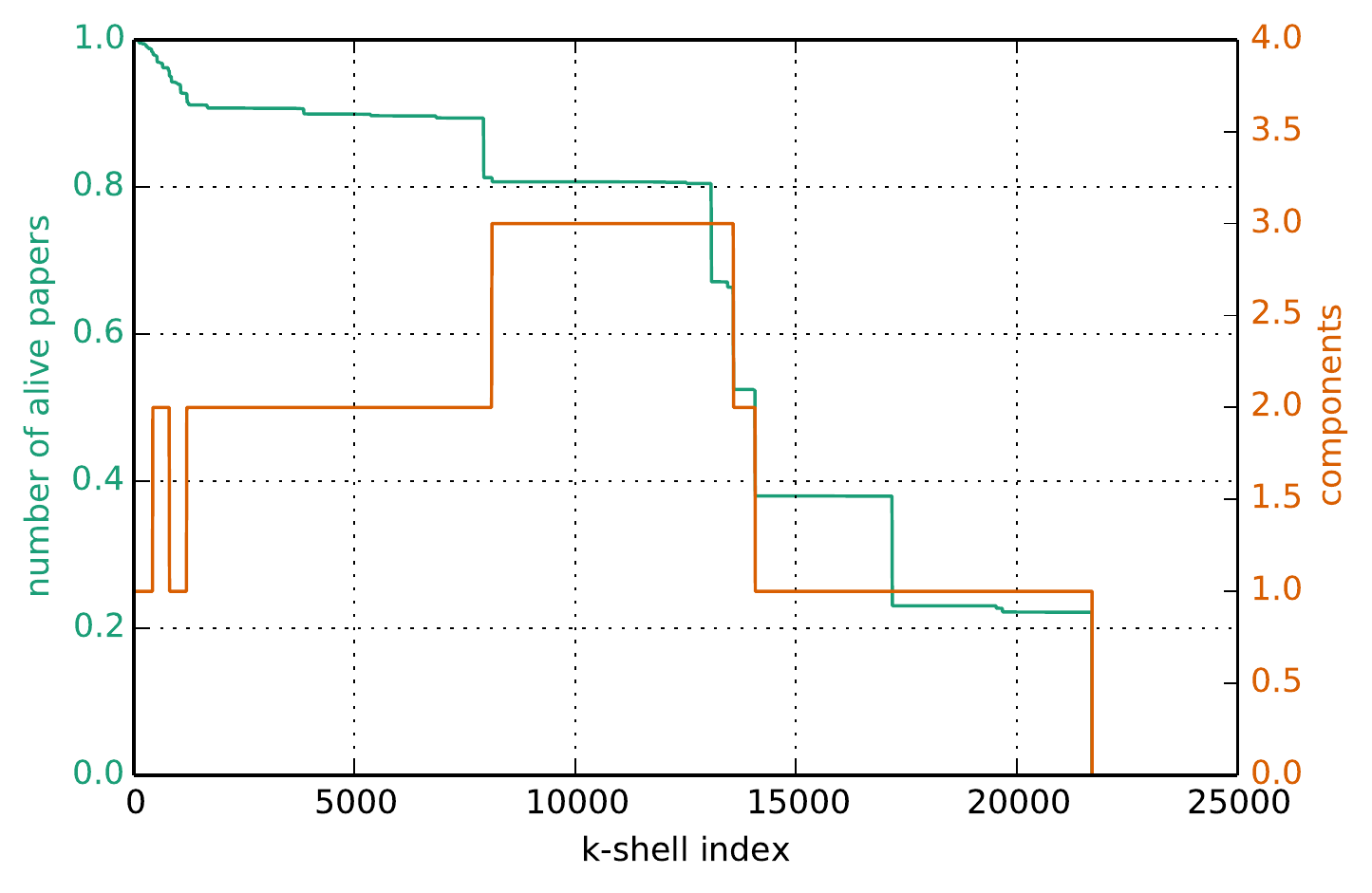}
  \caption{New York}
\end{minipage}
\caption{A comparison between a monocore and a multicore city. The alive papers at small $k$ are those still in the graph at the step $k$ of the k-shell algorithm.}
\label{fig:NYvsToronto}
\end{figure}

For different cities the behavior can be strikingly different, as we see in Figure \ref{fig:NYvsToronto}. In some cities, like Toronto, the dynamics are concentrated at the very beginning of the k-shell decomposition (low $k$ indices) and the fraction of the papers in the last shell is on the order of $1$. We shall characterize such cities as ``monocore''. On the other hand some cities, like New York, experience sizeable drops in the fraction of the alive papers through out the transition from $k=0$ to $k=k_{max}$ and the city is split in different components at various stages. We characterize these cities as ``multicore'' and each of the cities we discuss in the final subsection fall into this category.

We can further characterize the core structure of cities by observing how the city breaks down as $k$ increases. In that way we can define a functional ``core'' as a (large) set of publications that are geographically close to each other and represent a significant portion of the city's productivity. For the analysis below we define a core as such a group of publications with the same $k$-index that together are at least one tenth ($1/10$) the size (publications) of the $k_{max}$ core for that city.

Using the journal impact factor as a proxy for publication impact it is possible to estimate the quality of publications arising from each city and each core. In this way we can study the heterogeneity of research quality within cities. To do this we use the Thomson Reuters Journal Citation Report (JCR) Impact Factor for 2012. A journal's 2012 Impact Factor is the average number of citations accumulated in 2012 by a paper published in the previous two years (2010--2011). To estimate the average quality of research within a core we simply average the Impact Factors of each publication in that core. This quantity can be thought of as the expected number of citations an ``average'' paper from the given core will accumulate in its first two years. It is a coarse measure as the individual papers will receive more or less citations than the journal average and it is possible that the output of specific cities or cores changes significantly over time. However, we are averaging over thousands and tens of thousands of papers for each core so the average Impact Factor is a reasonable measure of impact for qualitative comparison. We have also tested with the 5-year Impact factor and Impact Factors from earlier years and there is no change in the qualitative results.

\section{Results}

\subsection{The structure of the leading clusters}

In this section we present results of our analysis for three major cities: London, Boston, and Tokyo. These cities are three of the largest according to the number of unique addresses discovered in our geocoding process and are among the top producers in terms of total publications. For each city, we present our results in three ways. First we show a descriptive map of core and k-shell structure of the city. In these maps each core is depicted as a polygon whose color is proportional to the number of papers published within the core. The color scale runs from red, to orange, to yellow, with the saturation point being approximately 42,000 papers as seen below in Boston. In addition to the polygonal cores, all other points that produced a publication are also indicated by small dot, with color proportional to the point's k-shell. Each core is also labeled with the core's rank in terms of total publications produced. The second way we present our results is in a table containing summary statistics for each core. Using the rank appearing in both the map and the table one can see a variety of properties for each core. These are the k-shell, total number of publications and their average Impact Factor as well as the main affiliation produced by our affiliation disambiguation approach. It is worth noting that the affiliations appearing in these tables have only been minimally altered from the direct output of our algorithm. Changes were only made in cases where the affiliation was simply too long, a good example being ``Royal Postgraduate Medical School'' in London has been abbreviated as ``RPMS''. The final figure we provide describing our city level results is a graph that tracks three quantities as a function of increasing k-shell. The first quantity is the total number of papers at that k-shell or higher. The second quantity is the number of distinct clusters present at a given k-shell. The third is the average Impact Factor of papers at that k-shell or higher. It is our conjecture that the evolution of these three curves can be used to better quantitatively characterize the structure of the cluster.

\begin{figure}[H]
\centering
\includegraphics[width=0.50\textwidth]{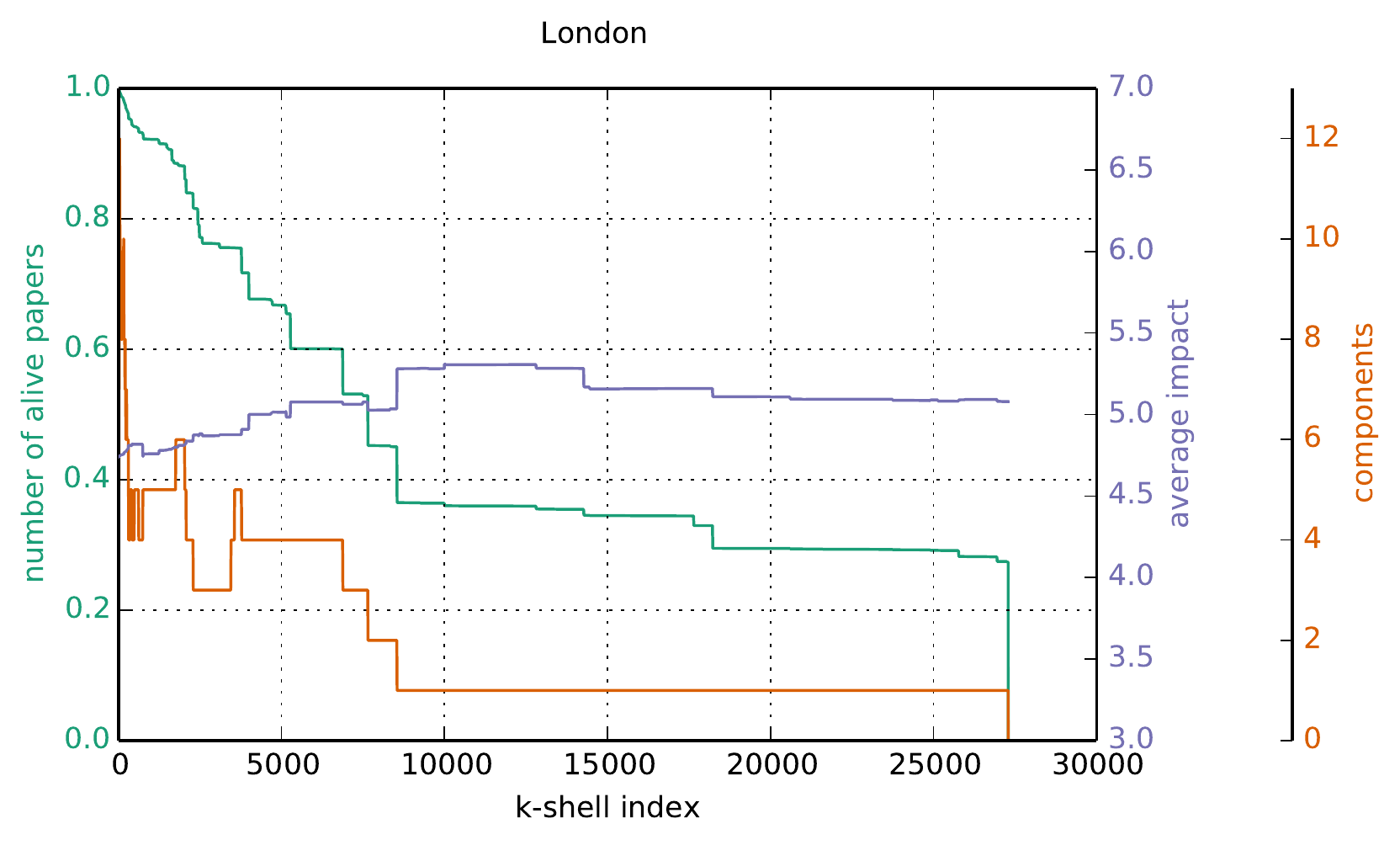}
\caption{London}
\label{fig:LondonGraph}
\end{figure}

Turning our attention to the London cluster found in Figure~\ref{fig:LondonGraph} we observe seven main cores. These cores seem to generally capture the geography of biomedical research within the city, hitting the main institutions. In comparison to the other top clusters London is particularly dominated by the first core, 27.3\% of the papers lie in the first core as opposed by 8.6\% in the second. While the most common affiliation (as determined by our affiliation disambiguation algorithm) in the first core is University College London, there is also a significant number of London School of Hygiene and Tropical Medicine, UCL Institute of Neurology, and University College Hospital affiliations. In reporting these affiliations, a limitation of the disambiguation algorithm is clear as UCL Institute of Neurology is indeed a part of UCL. However, given the general difficulty of institutional disambiguation the results are strong overall. In terms of IF a degree of heterogeneity is seen across London's cores. The large research Universities (UCL and Imperial College) display slightly higher average IF. The k-shell structure of London is largely dominated by the first core with the peripheral cores falling out quickly.

\begin{figure}[H]
\centering
\includegraphics[width=0.50\textwidth]{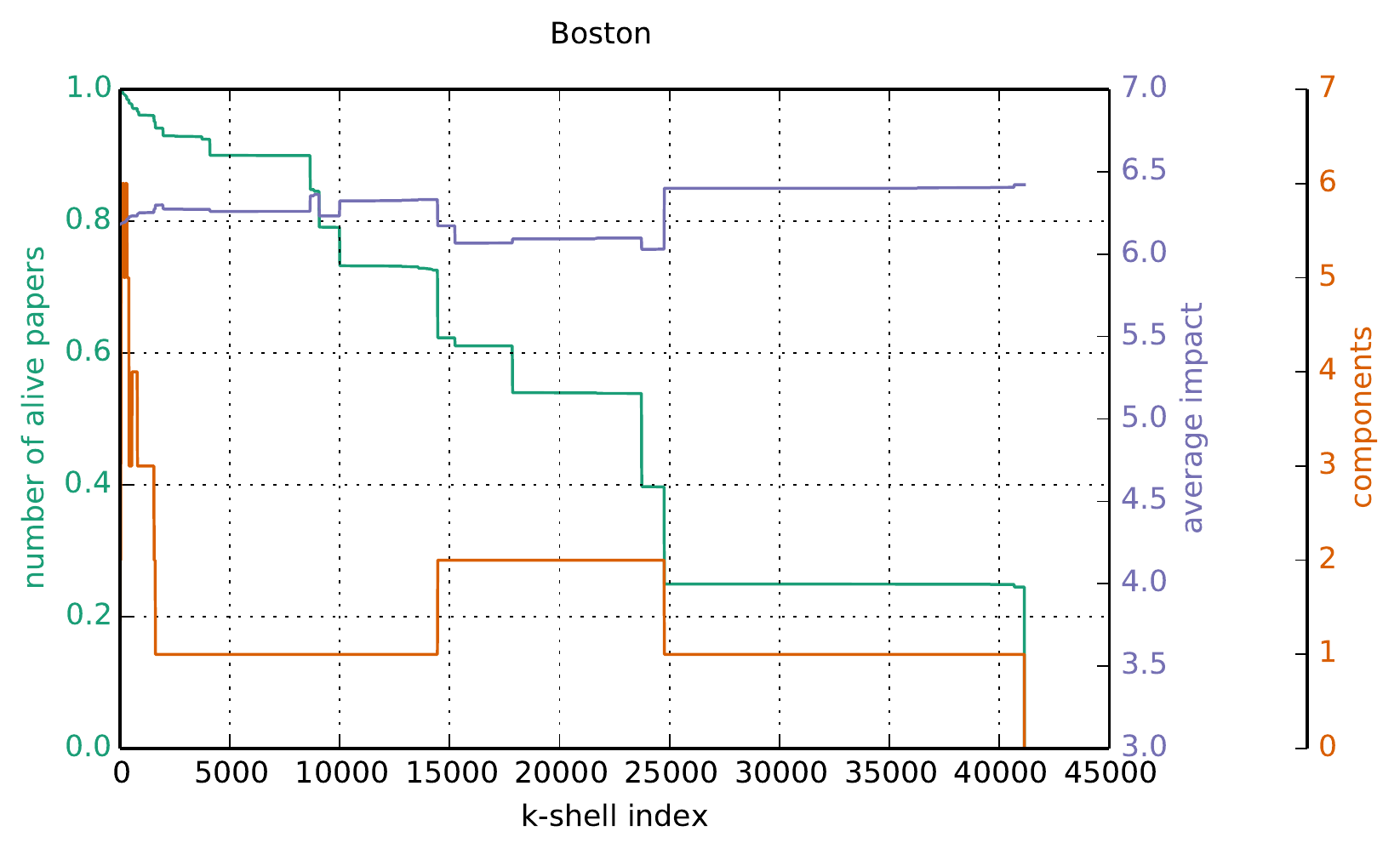}
\caption{Boston}
\label{fig:BostonGraph}
\end{figure}

The Boston cluster, found in Figure \ref{fig:BostonGraph}, spans two US counties, Suffolk and Middlesex and of all clusters identified in our analysis it is the largest in terms of papers produced. This cluster has been previously identified as top performing in other analysis~\cite{owen2002comparison}. The cores identified capture the main points of production within the area: Longwoods Medical Area, Massachusetts General Hospital, Massachusetts Institute of Technology (MIT), Harvard, etc. At the same time cores \#3 and \#5 represent a direction for further improvement of our geocoding, as some portion of papers originating in the Longwoods Medical Area were placed at the centroids of two zip codes (02215 and 02115). Aggregating those two with the first core we note again that the Longwoods Medical Area accounts for approximately half of the publications in the entire Boston region. Looking for additional affiliations within the first core we note, in addition to Harvard Medical School, Brigham and Women's Hospital, Beth Israel Deaconness Medical Center, Children's Hospital Boston, and Dana-Farber Cancer Institute. The second core is dominated by Massachusetts General Hospital. Core \#4 is largely MIT, but in jumping the Charles River also covers Boston University's main campus and that is reflected by the disambiguated affiliations returned by our algorithm. Interestingly, in core \#7 the presence of the Broad Institute was also detected. In terms of IF this cluster is somewhat heterogeneous, perhaps along lines of academic prestige. From the k-shell analysis the fact there are several significant drops in the curve indicates that Boston is, indeed, an area with many distinct centers of production.

\begin{figure}[H]
\centering
\includegraphics[width=0.50\textwidth]{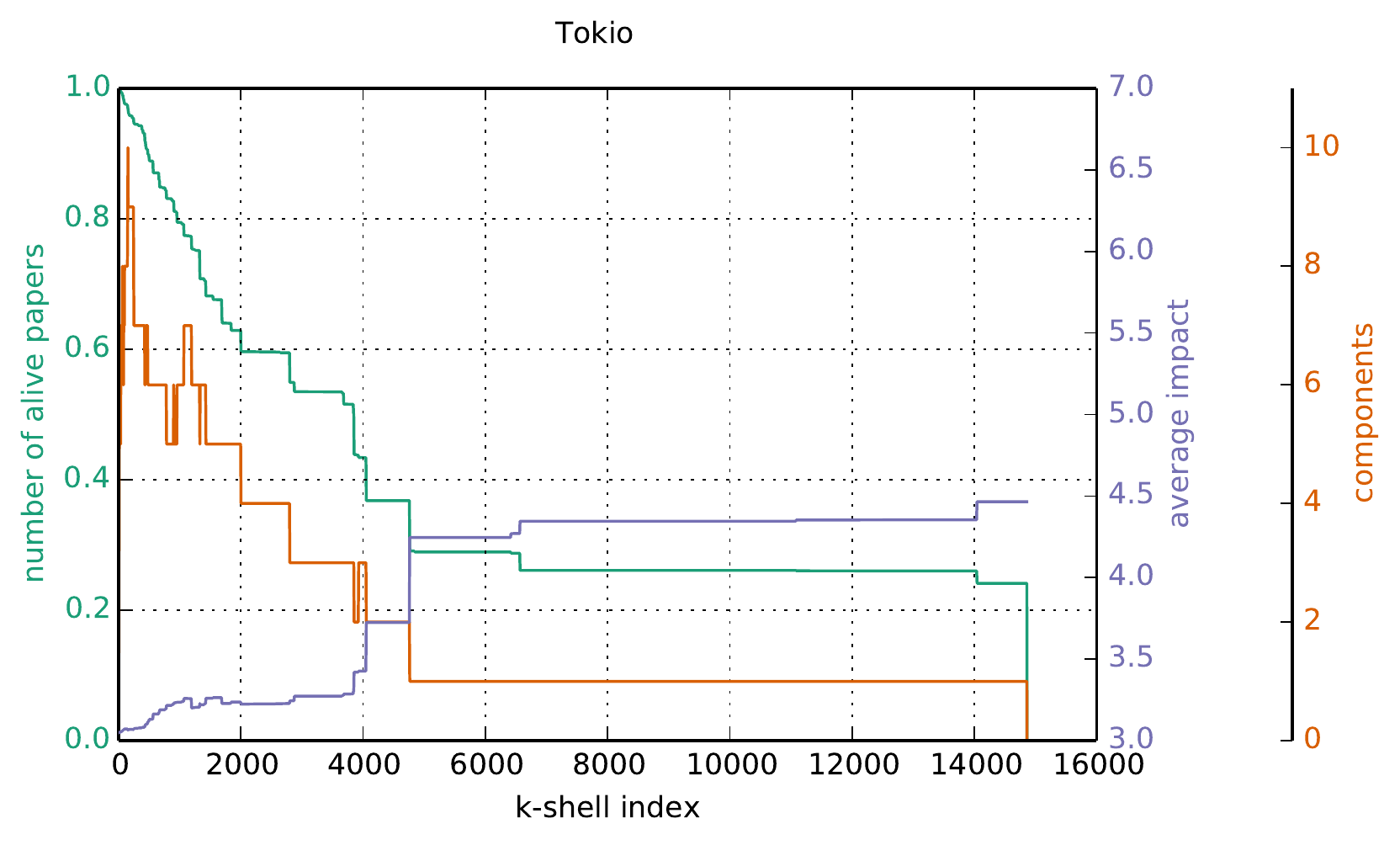}
\caption{Tokyo}
\label{fig:TokyoGraph}
\end{figure}

For the Tokyo cluster the algorithm seems to have hit the main loci of production, as shown in Figure~\ref{fig:TokyoGraph}. The first core corresponds to the University of Tokyo main campus. In the second core an additional significant affiliation produced by the disambiguation is Toranomon Hospital. Core \#5 is not an error, but actually the Komaba campus of the University of Tokyo. Within the fifth core the disambiguation algorithm also correctly identified Tokai University, which is present there though its Yoyogi campus. Within Tokyo we observe a large variation in both the productivity of cores, as well as their average IF. Indeed in the case of Tokyo the range of average IFs is very wide, but does not seem to correlate directly with productivity. The k-shell analysis correctly represents the city as one dominated by a single core, the University of Tokyo main campus.

\clearpage
\begin{figure}[H]
\centering
\includegraphics[width=1.00\textwidth]{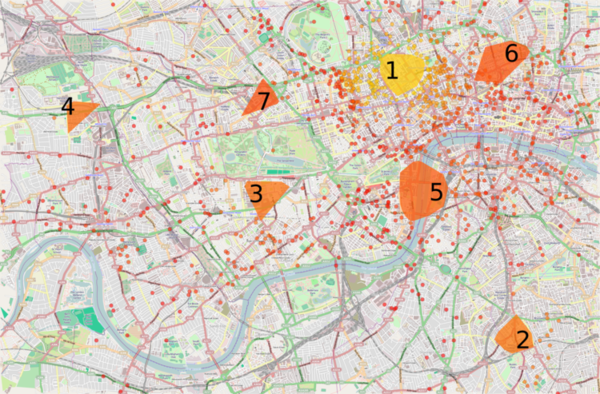}
\label{fig:LondonMap}
\end{figure}

\begin{center}
\begin{tabular}{c}
{\bf City 1: London} \\ \hline
	\begin{tabular}{cc}
	99729 Publications & Maximum K-Shell: 27288 \\
	Average IF: 4.74 & Core Threshold: 2729 Papers
	\end{tabular} \\ \hline
	{\tiny
	\begin{tabular}{cc}
		Core \#1 & Core \#2 \\ \hdashline
			\begin{tabular}{cc}
			Affiliation(s): & University College London \\
			Publications: & 27291 \\
			K-Shell: & 27288 \\
			Average IF: & 5.08 \\
			\end{tabular}
		&
			\begin{tabular}{cc}
			Affiliation(s): & King's College London \\
			Publications: & 8538 \\
			K-Shell: & 8537 \\
			Average IF: & 3.98 \\
			\end{tabular}
		\\
		Core \#3 & Core \#4 \\ \hdashline
			\begin{tabular}{cc}
			Affiliation(s): & Imperial College London \\
			Publications: & 7645 \\
			K-Shell: & 7644 \\
			Average IF: & 5.37 \\
			\end{tabular}
		&
			\begin{tabular}{cc}
			Affiliation(s): & RPMS, Hammersmith Hospital \\
			Publications: & 6874 \\
			K-Shell: & 6873 \\
			Average IF: & 5.19 \\
			\end{tabular}
		\\
		Core \#5 & Core \#6 \\ \hdashline
			\begin{tabular}{cc}
			Affiliation(s): & St Thomas' Hospital \\
			Publications: & 5329 \\
			K-Shell: & 5268 \\
			Average IF: & 3.93 \\
			\end{tabular}
		&
			\begin{tabular}{cc}
			Affiliation(s): & Moorfields Eye Hospital \\
			Publications: & 4000 \\
			K-Shell: & 3994 \\
			Average IF: & 3.38 \\
			\end{tabular}
		\\
		Core \#7 &  \\ \cdashline{1-1}
			\begin{tabular}{cc}
			Affiliation(s): & St Mary's Hospital \\
			Publications: & 3772 \\
			K-Shell: & 3769 \\
			Average IF: & 4.24 \\
			\end{tabular}
		&
	\end{tabular}
	}
\end{tabular}
\end{center}

\clearpage
\begin{figure}[H]
\centering
\includegraphics[width=0.90\textwidth]{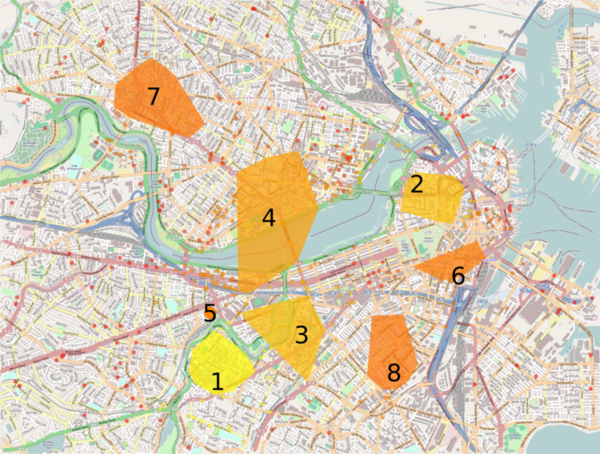}
\label{fig:BostonMap}
\end{figure}

\begin{center}
\begin{tabular}{c}
{\bf City 2: Boston} \\ \hline
	\begin{tabular}{cc}
	167958 Publications & Maximum K-Shell: 41138 \\
	Average IF: 6.18 & Core Threshold: 4114 Papers
	\end{tabular} \\ \hline
	{\tiny
	\begin{tabular}{cc}
		Core \#1 & Core \#2 \\ \hdashline
			\begin{tabular}{cc}
			Affiliation(s): & Harvard Medical School \\
			Publications: & 41140 \\
			K-Shell: & 41138 \\
			Average IF: & 6.42 \\
			\end{tabular}
		&
			\begin{tabular}{cc}
			Affiliation(s): & Massachusetts General Hospital \\
			Publications: & 24754 \\
			K-Shell: & 24752 \\
			Average IF: & 5.41 \\
			\end{tabular}
		\\
		Core \#3 & Core \#4 \\ \hdashline
			\begin{tabular}{cc}
			Affiliation(s): & Brigham and Women's Hospital \\
			Publications: & 23728 \\
			K-Shell: & 23723 \\
			Average IF: & 6.29 \\
			\end{tabular}
		&
			\begin{tabular}{cc}
			Affiliation(s): & Massachusetts Institute of Technology \\
			Publications: & 17236 \\
			K-Shell: & 14444 \\
			Average IF: & 7.30 \\
			\end{tabular}
		\\
		Core \#5 & Core \#6 \\ \hdashline
			\begin{tabular}{cc}
			Affiliation(s): & Beth israel Deaconness Medical Center \\
			Publications: & 11921 \\
			K-Shell: & 17848 \\
			Average IF: & 5.88 \\
			\end{tabular}
		&
			\begin{tabular}{cc}
			Affiliation(s): & Tufts University School of Medicine \\
			Publications: & 9785 \\
			K-Shell: & 9989 \\
			Average IF: & 5.09 \\
			\end{tabular}
		\\
		Core \#7 & Core \#8 \\ \hdashline
			\begin{tabular}{cc}
			Affiliation(s): & Harvard University \\
			Publications: & 9095 \\
			K-Shell: & 9056 \\
			Average IF: & 8.27 \\
			\end{tabular}
		&
			\begin{tabular}{cc}
			Affiliation(s): & Boston University School of Medicine \\
			Publications: & 8651 \\
			K-Shell: & 8648 \\
			Average IF: & 4.70 \\
			\end{tabular}
	\end{tabular}
	}
\end{tabular}
\end{center}

\clearpage
\begin{figure}[H]
\centering
\includegraphics[width=0.75\textwidth]{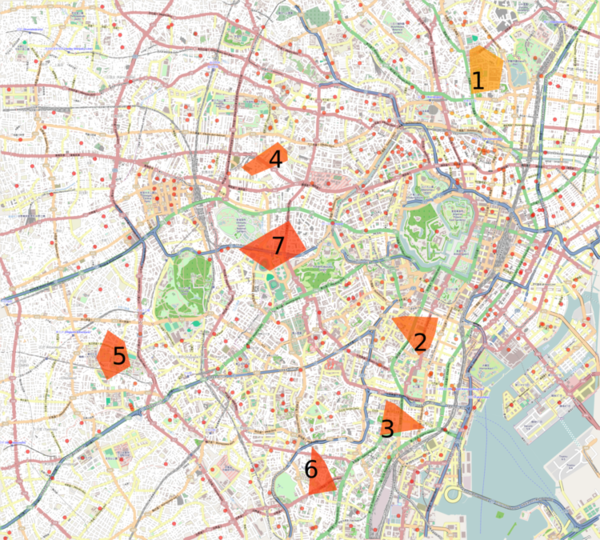}
\label{fig:TokyoMap}
\end{figure}

\begin{center}
\begin{tabular}{c}
{\bf City 3: Tokyo} \\ \hline
	\begin{tabular}{cc}
	61609 Publications & Maximum K-Shell: 14860 \\
	Average IF: 3.05 & Core Threshold: 1486 Papers
	\end{tabular} \\ \hline
	{\tiny
	\begin{tabular}{cc}
		Core \#1 & Core \#2 \\ \hdashline
			\begin{tabular}{cc}
			Affiliation(s): & University of Tokyo \\
			Publications: & 14861 \\
			K-Shell: & 14860 \\
			Average IF: & 4.47 \\
			\end{tabular}
		&
			\begin{tabular}{cc}
			Affiliation(s): & Jikei University School of Medicine \\
			Publications: & 4760 \\
			K-Shell: & 4759 \\
			Average IF: & 1.76 \\
			\end{tabular}
		\\
		Core \#3 & Core \#4 \\ \hdashline
			\begin{tabular}{cc}
			Affiliation(s): & Keio University \\
			Publications: & 4049 \\
			K-Shell: & 4048 \\
			Average IF: & 1.76 \\
			\end{tabular}
		&
			\begin{tabular}{cc}
			Affiliation(s): & Tokyo Women's Medical University \\
			Publications: & 3850 \\
			K-Shell: & 3849 \\
			Average IF: & 2.47 \\
			\end{tabular}
		\\
		Core \#5 & Core \#6 \\ \hdashline
			\begin{tabular}{cc}
			Affiliation(s): & University of Tokyo \\
			Publications: & 2798 \\
			K-Shell: & 2797 \\
			Average IF: & 3.03 \\
			\end{tabular}
		&
			\begin{tabular}{cc}
			Affiliation(s): & Kitasato University \\
			Publications: & 1999 \\
			K-Shell: & 1998 \\
			Average IF: & 3.48 \\
			\end{tabular}
		\\
		Core \#7 & \\ \cdashline{1-1}
			\begin{tabular}{cc}
			Affiliation(s): & Keio University School of Medicine \\
			Publications: & 1690 \\
			K-Shell: & 1686 \\
			Average IF: & 3.66 \\
			\end{tabular}
		&
		\\
	\end{tabular}
	}
\end{tabular}
\end{center}

\clearpage

\subsection{The dynamics of innovation clusters: the case of San Francisco}

\begin{figure}
\centering
\includegraphics[width=.7\textwidth]{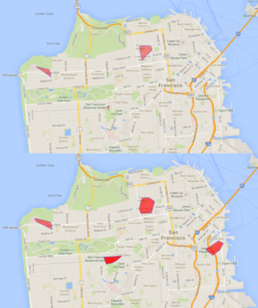}
\caption{The San Francisco cluster up to year 1998 (top) and from 1999 to present (bottom).}
\label{fig:SF}
\end{figure}

One of the main advantages of our approach it that is allows for the tracking of clusters and cores over time, in a variety of dimensions including size, shape, and content (specialization). To demonstrate this potential we consider the temporal evolution of the cluster centered on the city of San Francisco (light green in Figure~\ref{fig:bay-area}). In Figure~\ref{fig:SF} one can see the cores of biomedical productivity before, and after, the year 1998. The year 1998 was selected as the separation point since it was the year in which the Mission Bay area redevelopment project started. Up to 1998 the biomedical research activities in San Francisco were concentrated in the two main campuses of the University of California at San Francisco: the Mount Zion area, which stretches up to the California Pacific Medical Center and the Parnassus Heights campus followed by the San Francisco Veterans Affairs Medical Center. After 1998 the cluster extends and cores appear in the southern part of San Francisco. The Parnassus Medical Center underwent rapid development, becoming the main core of biomedical production immediately followed by the newly established Mission Bay campus. UCSF's Mission Bay Campus is the largest ongoing biomedical construction project in the world. Such an important change in the structure of the San Francisco biomedical area would have gone unnoticed by traditional approaches to the identification of innovation clusters. As illustrated by this example of San Francisco, our method allows for the tracking of the evolution of a cluster and its cores in space, ecology and technological content/specialization.

\section{Final discussion}

The Local agglomeration of innovative activities is at the center of attention for a widespread community of researcher in various disciplines, and the cluster idea is one of the most popular topics amongst policy makers and local economic practitioners. Despite its widespread appeal there is still a lack of consensus on the appropriate geographical entities for empirical analysis and policy intervention. 

In this paper we discussed some significant limitations of the common practice of relying on static and exogenously defined administrative regions to analyze innovation clusters, issues exacerbated in cross-country studies when regions are defined according to distinct national criteria. These limits are particularly restrictive when the emphasis is placed on the dynamic evolution of clusters over time. 

We provide a new methodology for endogenously identifying clusters, based on the location of activities (activity-based clustering) and apply it to the case of research institutions involved in scientific and technological production. Our method exploits a network approach to identify the relevant production clusters. In this respect, it allows for the identification of local concentrations of related inventive activities, offering some clear advantages in terms of precision, flexibility and applicability to the analysis of clusters in time, across countries and data sources. Moreover, our method is computationally efficient and relies upon publicly available data and software. To the best of our knowledge, ours is the only method which allows to identify clusters based on the existence of local bundles of related geo-referenced activities rather than assuming that knowledge interdependencies always exist within administrative borders. 

However our approach is not without limitations, that should be addressed in future work. First, geocoding should be improved, especially as far as country coverage is concerned, by testing alternative solutions for geo-referencing (i.e. Google places API or other similar services). Second, disambiguation can be refined by using repositories of names of individuals and institutions. Third, a better mathematical approach for characterizing the structure of clusters should be developed for the purposes of comparison across clusters and with random distributions of R\&D activities. Despite current limitations, we are confident that recent developments in open data policies and data-driven economic and geographical analyses will facilitate the development of better tools for dynamic content-based analysis of clusters. 

Further work in this area could move in a number of directions. First, different criteria can be used to draw links among nodes within a given distance. Namely, by using alternative data sources that provides multiple affiliations for patents and scientific papers, one could consider the number of co-authored papers as a proxy for real local relationships. In the current implementation of our method we assume that all researches active at a given distance are potentially connected. In future work the set of local connections could be restricted based on three criteria: (1) the analysis can be confined to the network made by actual links between co-located individuals such as collaborative or input-output relationships; (2) the set of potential collaboration at a given point in time can be limited by excluding individuals who are no more active at a given location and (3) only relationships among individuals working in related fields could be considered.  
Links between scientific and technological fields can be identified based on some population-based measure of relatedness of research areas within firms and in scientific and technological outputs. 
One more research opportunity is to analyze the evolution of clusters in time. This is essential to fully develop an evolutionary theory of cluster dynamics \cite{boschma2006} and to empirically test the predictions about cluster life cycles \cite{press2006,menzel2010}. 
Additionally, the relationship between clusters and productivity deserves further scrutiny. In the present work, we have identified a link between the average impact factor of scientific production and its location in the inner core of top biomedical clusters. This should be considered, however, only as preliminary evidence of the possible presence of locational advantages. More sound evidence should be produced in the future, based on a proper definition of innovative clusters. Another possibility for future work is to use our method to better investigate the dynamics of firm location decisions. Since we cluster both institution names and geographical areas, it is possible to dynamically investigate the impact of firm and individual mobility on regional performances. In this respect, one of the main advantages of our method is that it can be applied to multiple data sources on scientific publications, patents and the location of firms and other research organizations. This allows for the testing of  causal relationships between time-dependent location-specific advantages and firm performances. The range of new possibilities to better analyze the relationship between innovation networks and geographical clusters on the empirical ground can also contribute to the design of better policies and a more precise assessment of their effectiveness.

\section{Acknowledgements}
We thank F. Pammolli, A.M. Petersen, G. Morrison, A. Rungi, V. Tortolini, and R. Belderbos for helpful discussions. All authors acknowledge funding under the PNR ``Crisis
  Lab'' and SWITCH projects hosted by IMT. OP acknowledges funding from the Canadian Social Sciences and Humanities Research Council of Canada.

\section{References}
\bibliography{bibliography}{}

\begin{thebibliography}{10}

\bibitem{alcacer2012local}
Juan Alc{\'a}cer and Minyuan Zhao.
\newblock Local r\&d strategies and multilocation firms: The role of internal
  linkages.
\newblock {\em Management Science}, 58(4):734--753, 2012.

\bibitem{balland2013proximity}
Pierre-Alexandre Balland, Ron Boschma, and Koen Frenken.
\newblock Proximity and innovation: From statics to dynamics.
\newblock Technical report, Utrecht University, Section of Economic Geography,
  2013.

\bibitem{boschma2006}
Ron~A Boschma and Koen Frenken.
\newblock Why is economic geography not an evolutionary science? towards an
  evolutionary economic geography.
\newblock {\em Journal of Economic Geography}, 6:273--302, 2006.

\bibitem{carmi2007model}
Shai Carmi, Shlomo Havlin, Scott Kirkpatrick, Yuval Shavitt, and Eran Shir.
\newblock A model of internet topology using k-shell decomposition.
\newblock {\em Proceedings of the National Academy of Sciences},
  104(27):11150--11154, 2007.

\bibitem{cerina2013networks}
Federica Cerina, Alessandro Chessa, Fabio Pammolli, and Massimo Riccaboni.
\newblock Network communities within and across borders.
\newblock {\em Scientific Reports 2014}, (4):4546, 2014.

\bibitem{delgado2012}
Mercedes Delgado, Michael~E Porter, and Scott Stern.
\newblock Clusters, convergence, and economic performance.
\newblock Technical report, National Bureau of Economic Research, 2012.

\bibitem{delgado2013}
Mercedes Delgado, Michael~E Porter, and Scott Stern.
\newblock Defining clusters of related industries.
\newblock Technical report, mimeo, 2013.

\bibitem{diggle1991}
Peter~J Diggle and Amanda~G Chetwynd.
\newblock Second-order analysis of spatial clustering for inhomogeneous
  populations.
\newblock {\em Biometrics}, pages 1155--1163, 1991.

\bibitem{duranton2005}
Gilles Duranton and Henry~G Overman.
\newblock Testing for localization using micro-geographic data.
\newblock {\em The Review of Economic Studies}, 72(4):1077--1106, 2005.

\bibitem{feser2000}
Edward~J Feser and Stuart~H Sweeney.
\newblock A test for the coincident economic and spatial clustering of business
  enterprises.
\newblock {\em Journal of Geographical Systems}, 2(4):349--373, 2000.

\bibitem{furman2005public}
Jeffrey~L Furman, Margaret~K Kyle, Iain Cockburn, and Rebecca~M Henderson.
\newblock Public \& private spillovers, location and the productivity of
  pharmaceutical research.
\newblock {\em Annales d'{\'E}conomie et de Statistique}, (79/80):165--188,
  2005.

\bibitem{jacobs1970economy}
Jane Jacobs.
\newblock {\em The Death and Life of Great American cities}.
\newblock New York: Random House, 1961.

\bibitem{jonnalagadda2010nemo}
Siddhartha Jonnalagadda and Philip Topham.
\newblock Nemo: Extraction and normalization of organization names from pubmed
  affiliation strings.
\newblock {\em Journal of Biomedical Discovery and Collaboration}, 5:50, 2010.

\bibitem{lai2011disambiguation}
Ronald Lai, Alexander D'Amour, Amy Yu, Ye~Sun, Vetle Torvik, and Lee Fleming.
\newblock Disambiguation and co-authorship networks of the {US} patent inventor
  database.
\newblock {\em Harvard Institute for Quantitative Social Science, Cambridge,
  MA}, 2138, 2011.

\bibitem{marcon2003}
Eric Marcon and Florence Puech.
\newblock Evaluating the geographic concentration of industries using
  distance-based methods.
\newblock {\em Journal of Economic Geography}, 3(4):409--428, 2003.

\bibitem{marcon2009}
Eric Marcon and Florence Puech.
\newblock Measures of the geographic concentration of industries: improving
  distance-based methods.
\newblock {\em Journal of Economic Geography}, pages 1--18, 2009.

\bibitem{marshallprinciples}
Alfred Marshall.
\newblock {\em Principles of Economics}.
\newblock London: MacMillan, 1920.

\bibitem{martin2003deconstructing}
Ron Martin and Peter Sunley.
\newblock Deconstructing clusters: chaotic concept or policy panacea?
\newblock {\em Journal of economic geography}, 3(1):5--35, 2003.

\bibitem{maskell2006}
Peter Maskell and Le{\"\i}la Kebir.
\newblock What qualifies as a cluster theory?
\newblock {\em Clusters and regional development: Critical reflections and
  explorations}, page~30, 2006.

\bibitem{maskell2007}
Peter Maskell and Anders Malmberg.
\newblock Myopia, knowledge development and cluster evolution.
\newblock {\em Journal of Economic Geography}, 7:603--618, 2007.

\bibitem{menzel2010}
Max-Peter Menzel and Dirk Fornahl.
\newblock Cluster life cycles--dimensions and rationales of cluster evolution.
\newblock {\em Industrial and Corporate Change}, 19(1):205--238, 2010.

\bibitem{1990competitive}
Porter Michael.
\newblock The competitive advantage of nations.
\newblock {\em Harvard Business Review}, 68(2):73--93, 1990.

\bibitem{owen2002comparison}
Jason Owen-Smith, Massimo Riccaboni, Fabio Pammolli, and Walter~W Powell.
\newblock A comparison of us and european university-industry relations in the
  life sciences.
\newblock {\em Management science}, 48(1):24--43, 2002.

\bibitem{perroux1950economic}
Francois Perroux.
\newblock Economic space: theory and applications.
\newblock {\em The Quarterly Journal of Economics}, 64(1):89--104, 1950.

\bibitem{press2006}
Kerstin Press.
\newblock {\em A Life Cycle for Clusters?: The Dynamics of Agglomeration,
  Change, and Adaption}.
\newblock Springer, 2006.

\bibitem{rozenfeld2008laws}
Hern{\'a}n~D Rozenfeld, Diego Rybski, Jos{\'e}~S Andrade, Michael Batty,
  H~Eugene Stanley, and Hern{\'a}n~A Makse.
\newblock Laws of population growth.
\newblock {\em Proceedings of the National Academy of Sciences},
  105(48):18702--18707, 2008.

\bibitem{rozenfeld2009area}
Hern{\'a}n~D Rozenfeld, Diego Rybski, Xavier Gabaix, and Hern{\'a}n~A Makse.
\newblock The area and population of cities: New insights from a different
  perspective on cities.
\newblock {\em The American Economic Review}, 101(5):2205--25, 2011.

\bibitem{saxenian94}
A.~Saxenian.
\newblock {\em Regional Advantage. Culture and Competition in Silicon Valley
  and Route 128}.
\newblock Boston, MA: Harvard University Press, 1994.

\bibitem{ter2011co}
Anne~LJ Ter~Wal and Ron Boschma.
\newblock Co-evolution of firms, industries and networks in space.
\newblock {\em Regional Studies}, 45(7):919--933, 2011.

\bibitem{torvik2005probabilistic}
Vetle~I Torvik, Marc Weeber, Don~R Swanson, and Neil~R Smalheiser.
\newblock A probabilistic similarity metric for medline records: a model for
  author name disambiguation.
\newblock {\em Journal of the American Society for Information Science and
  Technology}, 56(2):140--158, 2005.

\end{thebibliography}
\bibliographystyle{plain}

\section{Appendix}

\subsection{Major biomedical clusters, US west cost}
\centering
\includegraphics[width=1\textwidth]{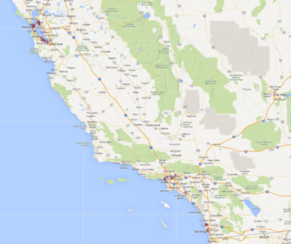}

\subsection{Major biomedical clusters, US east cost}
\centering
\includegraphics[width=1\textwidth]{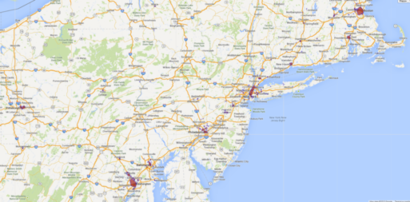}

\subsection{Major biomedical clusters in Northern Europe}
\centering
\includegraphics[width=1\textwidth]{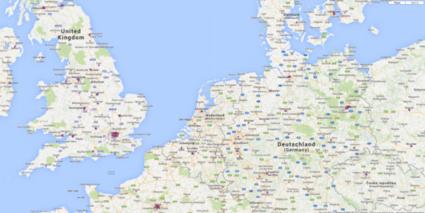}

\subsection{Main biomedical clusters in Japan}
\centering
\includegraphics[width=1\textwidth]{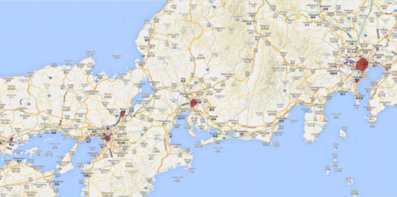}

\end{document}